\documentclass[%
 reprint,
 showpacs,preprintnumbers,
 bibnotes,
 amsmath,amssymb,
 aps,
 prb,
]{revtex4-1}

\usepackage{graphicx}
\usepackage{dcolumn}
\usepackage{bm}

\usepackage{color}
\usepackage{float}
\usepackage{amsfonts}

\begin{document}


\title{Electron spin separation without magnetic field}
\author{J. Paw\l{}owski}
\author{P. Szumniak}
\email{pawel.szumniak@gmail.com}
\author{A. Skubis}
\author{S. Bednarek}
\affiliation{
	Faculty of Physics and Applied Computer Science,
	AGH University of Science and Technology, Krak\'{o}w, Poland\\
}

\date{\today}

\begin{abstract}
A nanodevice capable of separating spins of two electrons confined in a quantum dot formed in a gated semiconductor nanowire is proposed. Two electrons confined initially in a single quantum dot in the singlet state are transformed into the system of two electrons confined in two spatially separated quantum dots with opposite spins. 
In order to separate the electrons' spins we exploit transitions between the singlet and the triplet state which are induced by resonantly oscillating Rashba spin-obit coupling strength. The proposed device is all electrically controlled and the electron spin separation can be realized within tens of picoseconds. The results are supported by solving numerically quasi-one-dimensional time-dependent Schroedinger equation for two electrons, where the electron-electron correlations are taken into account in the exact manner.

\vspace{1.35 cm}
\end{abstract}


\maketitle


\section{INTRODUCTION}
The prospect of building a quantum computer with entirely new computing capabilities compared to classical computers stimulates intensive research of physical phenomena that may be used to build its basic building blocks\cite{LADD}. Semiconductor nanostructures are a particular focus of interest since these systems can be directly integrated with classical electronics\cite{1,2}. The spin state of single electrons confined in semiconductor nanostructures emerged as a very promising candidate for encoding the quantum bit \cite{3, 4}. 
It relatively weakly interact with the surrounding environment which makes it robust to the decoherence. The main source of its decoherence is hyperfine interaction with nuclear spins. However with help of special prolonging techniques\cite{PROL} the coherence time of an electron spin qubit can reach even hundreds of microseconds \cite{ms} which is long enough to be able to perform an appropriate number of quantum logic operations executing a quantum algorithm.
Recent state of the art experiments showed that it is possible to confine electrons in electrostatic quantum dots and perform single-qubit operations \cite{5, 6, 7, 8},  two-qubit operations \cite{9, 10, 11, s13} as well as initialize and read out electron spin qubit. Also quantum dots defined in quasi one dimensional nanowires\cite{s1,s2a,s2b,s3,s4} seems to be very promising hosts for electron and hole spin qubits and even exotic particles such as Majorana fermions\cite{majoranas}.

In the first experimental implementation of quantum gates acting on the electron spin qubits, the energy levels of states with opposite spins were split in a magnetic field and coherent transitions between them were induced by absorbing microwaves with the energy equal to the Zeeman energy \cite{5, 6}. These experiments showed remarkable level of control over single electron spin qubits.
However, since one have to apply oscillating magnetic field which is difficult to be generated locally, such methods are not suitable for addressing individual spin qubits in multi-qubit quantum registers. Application of magnetic field causes continuous precession of spins of all the confined qubits in the register which makes it difficult to perform operations independently on single qubits without affecting state of neighbor spin qubits. Thus in order to fulfill scalability criterion for physical implementation of quantum computation the new techniques suitable for selective manipulation of spin qubits have to be developed. Several appealing proposal has been suggested in order to overcome this problem. The main solution is to manipulate individual spin qubits using electric fields which can be generated locally within the nanostructure.

One of the promising ways to realize electrically manipulation of spin qubits is to encode the qubit in the singlet and triplet states of two electron system\cite{st1}. In such an approach rotation of the singlet-triplet qubit on the Bloch sphere can be realized by electrically tunable exchange interaction or by exploiting gradients of the magnetic field. Recently manipulation of such singlet-triplet qubits was experimentally realized \cite{12,13,st2, st3,st4, st5, s14, s15, s16}.

Another approach to control spin qubits by electrical means is to take advantage of spin-orbit interaction which couples spin and orbital degrees of freedom. By inducing oscillating voltages one can ''shake" the electron with resonant frequency which in turns induces coherent oscillation of the spin - the  electric dipole spin resonance (EDSR) technique\cite{EDSR1, *EDSR2, *EDSR3}. Recently much attention has been put on the theoretical investigation\cite{s11,s12,s17,swojcik} as well as on the experimental implementation\cite{s1,23,s4,s18} of the EDSR techniques for controlling electron and hole spin qubits confined in nanowire quantum dots. Though still static external magnetic field is needed in this experimental setups.

The particularly promising experimental realization of fully magnetic-free control of electron spin was demonstrated very recently\cite{s19, *s20}, where the single electron spin rotations were realized by transporting the electron by surface acoustic waves in the presence of spin-orbit interaction. 

Furthermore several original techniques has been proposed in order to control electron and hole spin qubits without magnetic field. Single-qubit quantum logic operations can be realized without magnetic field by transporting an electron or hole along a two-dimensional closed path. A computer simulation of such nanodevices can be found in previous works by the authors \cite{14,*15,*HH1, *HH2}. It seems that the most difficult operation to perform without using a magnetic field is the spin initialization and spin readout. However, it is not utterly impossible. A nanodevice performing such an operation has been proposed in the Ref.~[\onlinecite{16}]. Such a device exploits phenomena of spin-dependent electron trajectory caused by spin-orbit interaction in order to distinguish electrons with different spins and transport them to separated parts of the nanodevice.
Another interesting method of spatial separation of electrons' spins density without using a magnetic field has been proposed in the work~[\onlinecite{17}], whose authors discuss a system of two electrons confined in a vertically coupled cylindrical double quantum dot. To perform operations on spins, they are using the Dzyaloshinskii-Moriya interaction \cite{18, *19, s6a, *s6b} controlled by eight lateral electrodes.

In this work, we propose novel technique and present nanodevice that is capable of separating the spin density of two electrons without using a magnetic field. Thus it allows for high fidelity electron spin initialization in either spin ''up" or spin ''down" state. The proposed nanodevice is characterize by several advantages comparing to previous proposals: it has a less complicated structure than nanodevice described in the Ref.~[\onlinecite{16}] and works much faster than the one suggested in the work~[\onlinecite{17}]. To separate the electron spin, we are employing the spin-orbit interaction with an oscillating in time coupling strength. This is possible in case of the Rashba spin-orbit (RSO) interaction, which strength can be modulated by an electric field generated locally by alternating voltage applied to the control electrodes. The RSO interaction with a variable amplitude has been used for various purposes in previous works, e.g. to study the oscillation of spin polarization in quantum rings \cite{20,s7,s8}, quantum wires \cite{21}, or in graphene \cite{22}, as well as to achieve complex spin and position dynamics of trapped cold fermionic atom\cite{s21}. In the present work interplay between resonantly oscillating RSO coupling strength and exchange interaction is used to generate transitions between singlet and triplet state, which are exploited further to generate two electron spin state where spin degrees of freedom are separated in different parts of the proposed nanodevice. The spin separation process is realized in ultrafast manner without use of magnetic field.

\section{DEVICE AND CALCULATION METHOD}
\begin{figure}[ht!]
\hspace{-7mm}
\includegraphics[width=6cm]{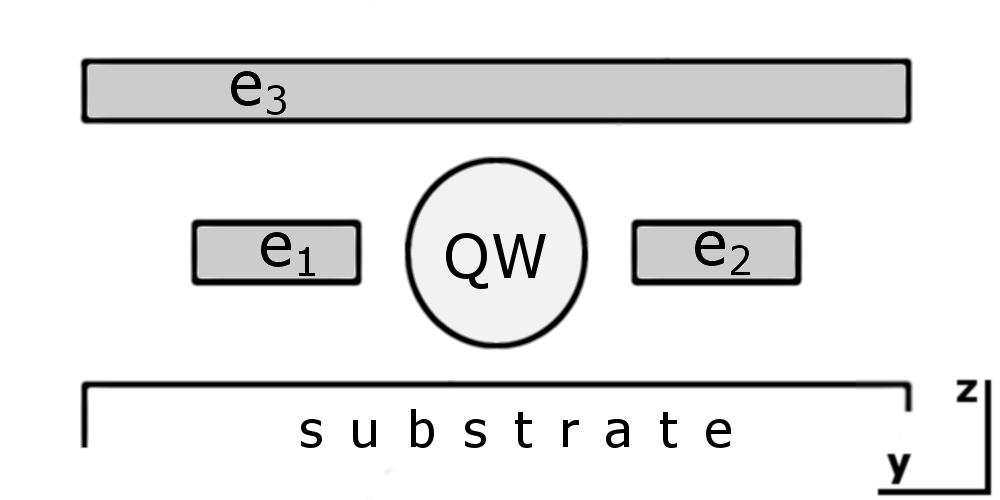}\\
\vspace{1.2cm}
\includegraphics[width=7.5cm]{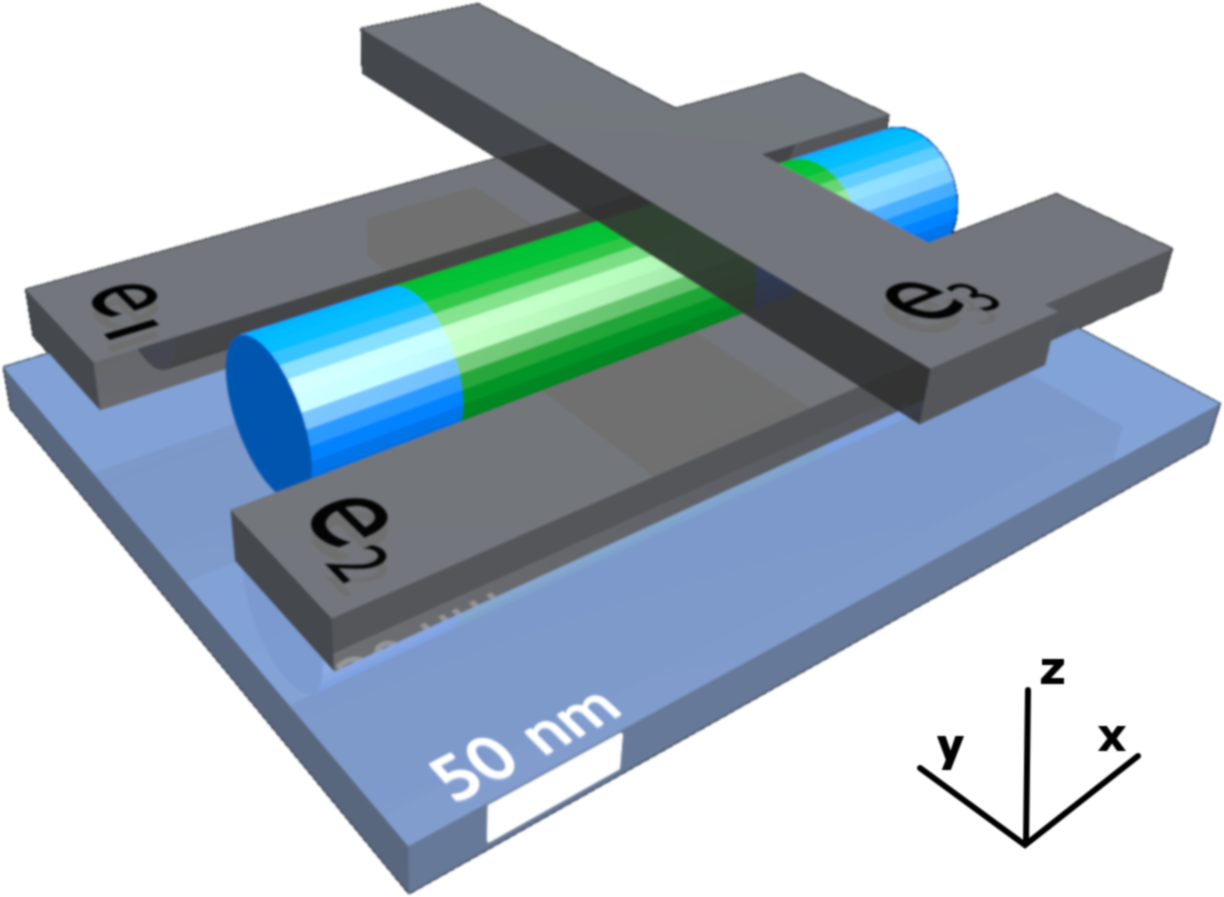}
\caption{\label{fig:1} (color online). Cross-section of the nanodevice (top). 
Schematic layout of electrodes \(\mathrm{e}_1\), \(\mathrm{e}_2\) and \(\mathrm{e}_3\) with respect to the quantum wire (QW) (bottom). }
\end{figure}

\begin{figure}
\includegraphics[width=6cm]{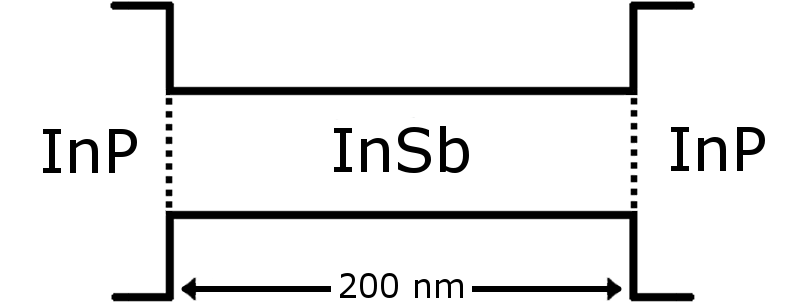}
\caption{\label{fig:2} 
A schematic band structure containing bottom conduction and top valence bands for InSb nanowire (green section of NW in Fig.~\ref{fig:1}) with InP ends (blue section of NW in Fig.~\ref{fig:1}).
}
\end{figure}

Let us consider a system of two electrons confined in a semiconductor nanostructure as depicted in Fig.~\ref{fig:1}. 
The proposed nanodevice is composed from the gated semiconductor quasi-one-dimensional quantum nanowire in similar manner as in the experimental setup from Ref. [\onlinecite{23}]. However in our calculations we use a section of the quantum wire with a length limited from two sides by a material that forms a potential barrier for the conduction band electrons (Fig.~\ref{fig:2}). Thus this part of a quantum wire (green section of the nanowire in Fig.~\ref{fig:1}) can be considered as a quantum dot in which electrons are confined.

The nanowire is placed between electrodes $e_1$ and $e_2$. By applying the alternating voltage to these electrodes one can induce oscillating in time RSO coupling strength. The electrode $e_3$ is placed above the center of the quantum wire and its role is to generate electrically tunable tunnel barrier which allows to switch the system between single and double quantum dot regime by electrical means.

\begin{figure}[ht!]
\includegraphics[width=7.5cm]{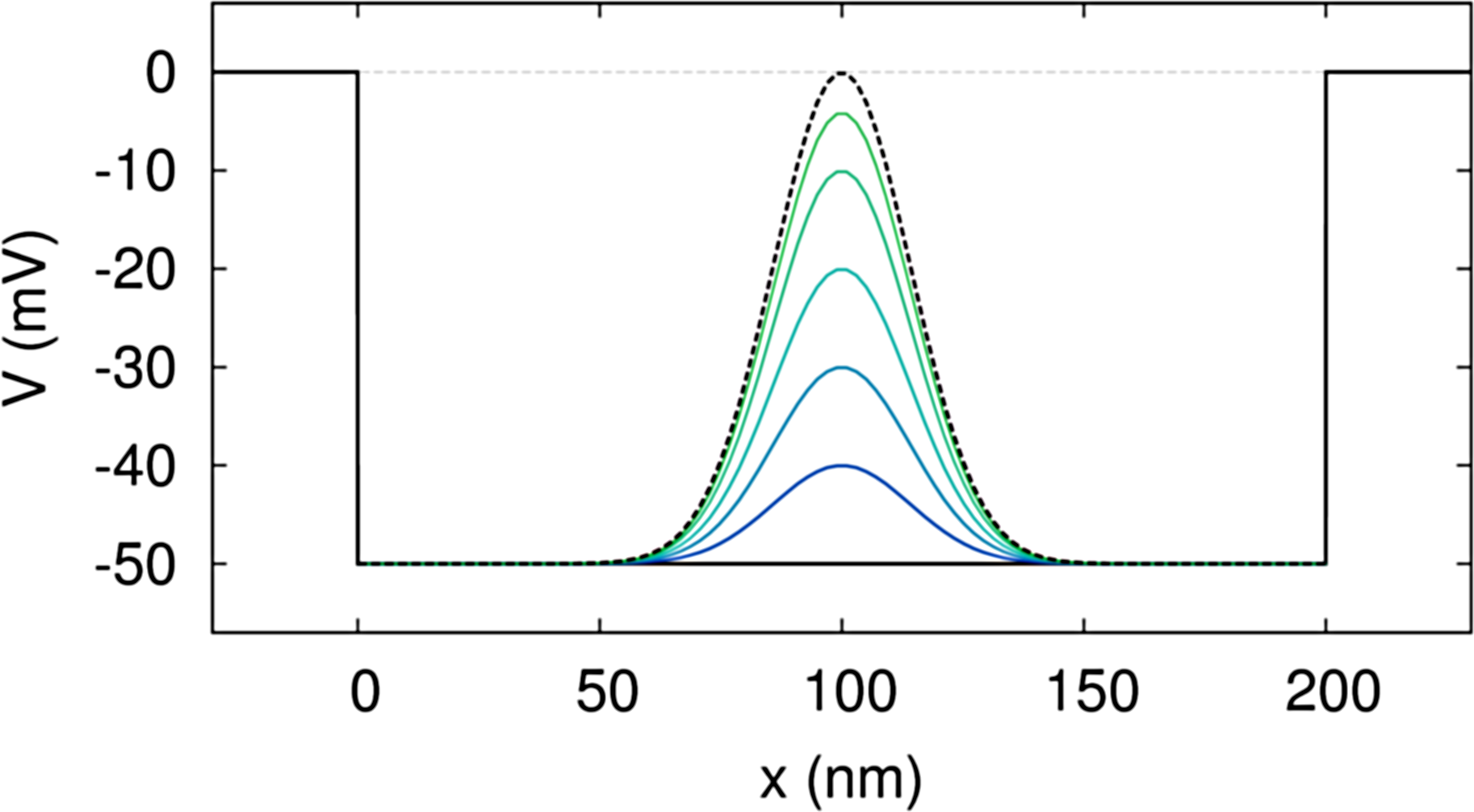}
\caption{\label{fig:3} (color online). The shape of the confinement potential along the quantum wire for various barrier heights genarated by the voltage applied to the electrode $e_3$.}
\end{figure}

In the quasi-one-dimensional limit the electrons confined in the considered nanostructure can be described by the following Hamiltonian:
\begin{equation}
H=\left(h^1_0+h^2_0+V_\mathrm{eff}(x_1\!-\!x_2)\right)\!1_4 +H^{SO},
\label{eq:8}
\end{equation}
with the single-electron energy operator:
\begin{eqnarray}
h^j_0=-\frac{\hbar^2}{2m^\ast}\frac{\partial^2}{\partial x^2_j}+|e|V\!(x_j,t),
\label{eq:9}
\end{eqnarray}
where $1_4$ is $4\times 4$ identity matrix, $m^*=0.014m_{0}$ is the effective mass of the electron in InSb nanowire, $m_{0}$ is the free electron mass and $x_j$ denotes the position of the j-th electron, $j=1,2$.
The voltage applied to electrode \(\mathrm{e}_3\) is the source of potential along the quantum wire - the interdot tunnel barrier, which we model with the following expression:
\begin{eqnarray}
V(x,t)=-V_0+V_1(t)\exp\!\left(-(x-x_0)^2/b^2\right),
\label{eq:1}
\end{eqnarray}
where we denote \(x_0=l/2\) as the midpoint of the quantum wire, $l=200$ nm is the length of the InSb nanowire section, \(V_0\) is the height of the potential barrier at the ends of the InSb section. The amplitude \(V_1(t)\) of the interdot potential barrier is controlled by voltage applied to electrode \(\mathrm{e}_3\). By changing the height \(V_1(t)\) of the barrier, we can tune the system from single quantum dot (\(V_1(t_{single})=0\)) to double quantum dot (\(V_1(t_{double})=V_0\)) regime as illustrated in Fig.~\ref{fig:3}.


We assume that the length of quantum wire (parallel to the \(x\)-direction) is much larger than its thickness and that confinement in the directions perpendicular to the quantum wire (\(y\) and \(z\)) is so strong that the spacing between energy levels due to lateral movement is much larger than the interaction energy of two electrons. Thus we can assume the approximate form of the wave function, which can be separated into parts responsible for the confinement in the directions perpendicular to the wire (\(y\), \(z\)) and a part describing the dynamics of motion in the direction parallel to the wire (\(x\)). A strong parabolic confinement in the \(y\) and \(z\) direction with cylindrical symmetry freezes the lateral wave functions of both electrons into the Gaussians:
\begin{eqnarray}
\Phi(y,\!z)=\left(\sqrt{\pi}d\right)^{-1}\!\exp\!\left(-\!\left( y^2+z^2\right)\!/2d^2\right),
\label{eq:2}
\end{eqnarray}
where the Gaussian dispersion parameter \(d\) = 25~nm determines the approximate radius of the nanowire. Upon integration in the directions perpendicular to the wire $\langle\Phi(y,z)|\frac{e^2}{4\pi\varepsilon_0\varepsilon|{\bf r_{12}}|}|\Phi(y,z)\rangle$, the electron-electron Coulomb interaction can be replaced by the effective interaction in one dimension\cite{24}:
\begin{eqnarray}
V_\mathrm{eff}(x_1\!-\!x_2)=\frac{e^2\sqrt{\pi/2}}{4\pi\varepsilon_0\varepsilon d}\,\mathrm{erfce}\!\left(\frac{|x_1\!-\!x_2|}{\sqrt{2}d}\right),
\label{eq:3}
\end{eqnarray}
where \(\mathrm{erfce}(x)\equiv\exp\!⁡(x^2)\left(1-\mathrm{erf}⁡(x)\right)\) and \(\mathrm{erf}⁡(x)\) is a standard error function. The dielectric constant is denoted by $\varepsilon$ and its value $\varepsilon=16.4$ is taken for InSb material. 

A similar approach leading to the dimensionality reduction of the system has been successfully applied in other work~[\onlinecite{30}], where authors explained appearance of the fractional resonances lines in experiments related with the EDSR in the nanowire quantum dots \cite{23, 23a}.

The voltage applied between the electrodes \(\mathrm{e}_1\) and \(\mathrm{e}_2\) generates an electric field parallel to the \(y\) axis, and thus perpendicular to the quantum wire axis. Consequently the RSO interaction is electrically generated (the main spin-orbit interaction type in the [111] grown InSb nanowires \cite{23}). In the case of frozen electron motion in transverse directions ($y,z$) RSO interaction can be described by the following Hamiltonian \cite{25,26}:
\begin{eqnarray}
H^j_{SO} = -\alpha k^j_x \sigma_z =
-\alpha\!\begin{pmatrix}
  k^j_x & 0 \\
  0 & - k^j_x
 \end{pmatrix},
\label{eq:4}
\end{eqnarray}
where \(\sigma_z\) is the \(z\)-component of the Pauli matrices, and the operator \(\hbar k^j_x\) is the momentum operator of the \(j\)-th electron confined in the wire aligned along the \(x\)-direction:
\begin{equation*}
  k^j_x =- i \frac{\partial}{\partial x_j} \equiv - i \partial_{x_j}.
\end{equation*}
The RSO coupling constant $\alpha$ will be modulated in time: $\alpha\rightarrow\alpha(t)$ corresponding to periodic changes of the voltage applied to the electrodes $e_1$ and $e_2$.\cite{comm}

In order to describe two electrons together with their orbital and spin degrees of freedom, we use following four-row spinor representation\cite{27} of the two-electron wave function:

\begin{eqnarray}
\Psi(x_1,x_2,t)=\!\begin{pmatrix}
 \psi_{\uparrow\uparrow}(x_1,x_2,t) \\
 \psi_{\uparrow\downarrow}(x_1,x_2,t) \\
 \psi_{\downarrow\uparrow}(x_1,x_2,t) \\
 \psi_{\downarrow\downarrow}(x_1,x_2,t)
 \end{pmatrix}.
\label{eq:5}
\end{eqnarray}
The arrows indicate the spin projection onto the quantization axis (\(z\)) of the first and second electron, respectively. Since we are dealing with fermions the total wave function~(\ref{eq:5}) has to be antisymmetric with respect to simultaneous exchange of the space and spin coordinates of both electrons. This symmetry property imposes the following constraints on the basis wave functions: \( \psi_{\uparrow\uparrow}(x_1,x_2)=-\psi_{\uparrow\uparrow}(x_2,x_1)\), \( \psi_{\downarrow\downarrow}(x_1,x_2)=- \psi_{\downarrow\downarrow}(x_2,x_1)\) and \( \psi_{\uparrow\downarrow}(x_1,x_2)=- \psi_{\downarrow\uparrow}(x_2,x_1)\). This can be written in more compact form using the SWAP matrix \(S\): \(\Psi(x_1,x_2)=-S \Psi(x_2,x_1)\). In the chosen representation, the spin operators for first and second electron have the following forms: \(\boldsymbol{\sigma}\otimes 1_2\), \(1_2 \otimes \boldsymbol{\sigma}\), where the Pauli vector is defined by Pauli matrices: \(\boldsymbol{\sigma}\equiv[\sigma_x, \sigma_y, \sigma_z]\), and \(1_2\) means the 2\(\times\)2 identity matrix. By using these operators, we can introduce the spin-orbit interaction operators for the first and second electron: \(H^1_{SO}\otimes 1_2\) and \(1_2 \otimes H^2_{SO}\). Since the \(z\)-component of the spin operator for the first and the second electron correspond to matrices:
\begin{eqnarray}
\hspace{-6mm}
\sigma_z\!\otimes\!1_2\!=\!\begin{pmatrix}
1 & 0 & 0 & 0\\
0 & 1 & 0 & 0\\
0 & 0 & -1 & 0\\
0 & 0 & 0 & -1\\
\end{pmatrix}\!\!,
\hspace{1mm}
1_2\!\otimes\!\sigma_z\!=\!\begin{pmatrix}
1 & 0 & 0 & 0\\
0 & -1 & 0 & 0\\
0 & 0 & 1 & 0\\
0 & 0 & 0 & -1\\
\end{pmatrix}\!\!,
\label{eq:6}
\end{eqnarray}
the spin-orbit part of the total Hamiltonian (1) is expressed by a diagonal matrix:
\begin{align}
& H^{SO} = H^1_{SO}\otimes 1_2 + 1_2 \otimes H^2_{SO} = i \alpha(t) \times \nonumber \\
&\times \mathrm{diag} \left(\partial_{x_1}\!\!+\!\partial_{x_2}, \partial_{x_1}\!\!-\!\partial_{x_2}, -\partial_{x_1}\!\!+\!\partial_{x_2}, -\partial_{x_1}\!\!-\!\partial_{x_2}\right).
\label{eq:7}
\end{align}

Stationary states e.g. the ground and the first excited state of the system can be found by solving the eigenvalue equation $H\Psi(x_1, x_2, t_0)=E\Psi(x_1, x_2, t_0)$ of the Hamiltonian~(\ref{eq:8}) with frozen time dependencies of all the parameters involved in the expressions~(\ref{eq:1}): $V_1(t_0)=const$ and in ~(\ref{eq:7}): $\alpha(t_0)=const$. In order to solve stationary Schroedinger equation we use image time propagation technique (ITP)\cite{ITP}. The time evolution of the system is described by the solution of the time dependent Schroedinger equation $i\hbar\frac{\partial}{\partial t}\Psi(x_1, x_2, t)=H\Psi(x_1, x_2, t)$  with Hamiltonian~(\ref{eq:8}) which is solved numerically by using explicit Askar-Cakmak scheme\cite{AC}:
\begin{align}
\Psi(x_1,x_2,t+dt)=&\,\Psi(x_1,x_2,t-dt)- \nonumber \\
&-2\frac{i}{\hbar}H(x_1,x_2,t)\Psi(x_1,x_2,t)dt.
\label{eq:10}
\end{align}
By using this approach with the wave function representation containing spin and orbital degrees of freedom as well as the effective Coulomb interaction we are able to take into account the electron-electron correlations in our model in the exact manner.

\section{RESULTS AND DISCUSSION}
\subsection{Stationary states}

Let us assume that the spin-orbit coupling constant is equal to zero \(\alpha(t_0)=0\) and that the interdot barrier is absent \(V_1(t_0)=0\). Both electrons will then be inside a single potential well. The ground state of such a system is the singlet state \(\left| S\right\rangle\), with antiparallel oriented spins.  In this state, the orbital part (symmetric) and spin part (antisymmetric) of the wave function are separable:
\begin{eqnarray}
\left| S\right\rangle=\varphi^S(x_1,x_2)\!\begin{pmatrix}
0 \\
1 \\
-1 \\
0
 \end{pmatrix}.
\label{eq:11}
\end{eqnarray}
The first excited state is the triplet state \(\left| T_0\right\rangle\), also with separable spatial (antisymmetric) and spin (symmetric) parts:
\begin{eqnarray}
\left| T_0\right\rangle=\varphi^{T}(x_1,x_2)\!\begin{pmatrix}
{\ }0{\ }\\
1 \\
1 \\
0
\end{pmatrix}.
\label{eq:12}
\end{eqnarray}
The spatial part of the singlet wave function \(\varphi^{S}(x_1,x_2)\) is presented on the left part of the Fig. 4, while the triplet wave function \(\varphi^{T}(x_1,x_2)\) is depicted on the right part of the Fig. 4. The energies of both energy levels are moved away due to exchange interaction.
\begin{figure}[ht!]
\includegraphics[width=8.7cm]{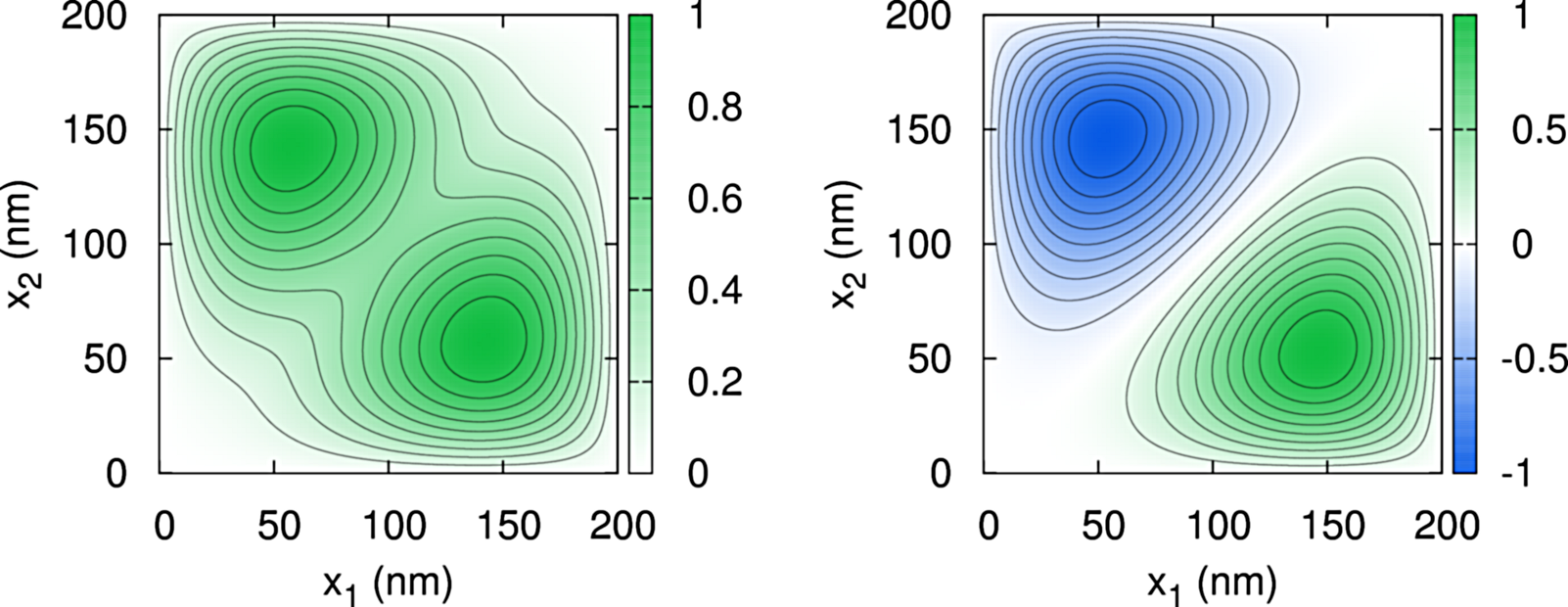}
\caption{\label{fig:4} (color online). The spatial part of the wave function of the ground state, singlet --- \(\varphi^{S}(x_1,x_2)\)  (left side) and the first excited state, triplet --- \(\varphi^{T}(x_1,x_2)\) (right side).}
\end{figure}
For the applied InSb material parameters (by solving stationary Schroedinger Equation with the ITP method) we get the singlet-triplet splitting energy \(E_{ST}=E_S-E_T\) equal to \(E_{ST}=0.936\)~meV. In the experimental setup one may tune the singlet-triplet splitting energy e.g. by keeping non zero height of the interdot tunnel barrier or simply by taking longer section of the InSb in the nanowire.

Let us consider another situation where the two electrons are separated by the interdot barrier $V_1=V_0$. If we start raising the potential barrier~(\ref{eq:1}) $V_1=0\rightarrow V_1=V_0$, the electrons will move away from each other and the exchange energy will decrease and eventually vanish. In our calculations we have assumed that \(V_0 = 50\)~mV and \(b = 20\)~nm in expression~(3). These parameters have been chosen so that, when $V_1 = V_0$, the singlet-triplet energy splitting is close to $E_{ST}=0.$\cite{comm0} Thus orbital part of the wave functions of the two-electron system can be constructed based on the single-electron functions localized completely in the left \(\varphi_{L}(x)\) and right \(\varphi_{R}(x)\)  part of the quantum wire:
\begin{align}
&\varphi^{S\,(T)}(x_1,x_2)= \nonumber \\
= &\frac{1}{\sqrt{2}}\!\left(\varphi_{L}(x_1)\,\varphi_{R}(x_2) +\!(-) \,\varphi_{R}(x_1)\,\varphi_{L}(x_2) \right).
\label{eq:13}
\end{align}
The two lowest energy states are the singlet and the triplet, but since functions \(\varphi_{L}\) and \(\varphi_{R}\) do not overlap, the exchange interaction disappears. Thus the singlet and triplet state are degenerate.

Let us note that if we construct a balanced linear combination of these wave functions (by adding together the overall spin-orbital wave functions), we will get:
\begin{eqnarray}
\left| \Psi^\mathrm{sep}_{+}\right\rangle=\!\frac{1}{\sqrt{2}}\!\left( \left|S\right\rangle\!+\!\left| T_0\right\rangle \right)=\!\begin{pmatrix}
0\\
\varphi_{L}(x_1)\,\varphi_{R}(x_2) \\
-\varphi_{R}(x_1)\,\varphi_{L}(x_2) \\
0
\end{pmatrix}\!.
\label{eq:14}
\end{eqnarray}
Comparing this function with the expression~(\ref{eq:5}), which defines representation, gives: \(\psi_{\uparrow\downarrow}(x_1,x_2)=\varphi_{L}(x_1)\,\varphi_{R}(x_2)\)  and \(\psi_{\downarrow\uparrow}(x_1,x_2)=-\varphi_{R}(x_1)\,\varphi_{L}(x_2)\).  It means that we have reached a state, in which the spin in the left dot is oriented in up ($+\hbar/2$) and in the right one spin is pointing down ($-\hbar/2$). 

On the other hand one can obtain analogous spin separated state:
\begin{eqnarray}
\left| \Psi^\mathrm{sep}_{-}\right\rangle=\!\frac{1}{\sqrt{2}}\!\left( \left|S\right\rangle\!-\!\left| T_0\right\rangle \right)=\!\begin{pmatrix}
0\\
\varphi_{R}(x_1)\,\varphi_{L}(x_2) \\
-\varphi_{L}(x_1)\,\varphi_{R}(x_2) \\
0
\end{pmatrix}\!,
\label{eq:14a}
\end{eqnarray}
which corresponds to the complementary situation where the spin in the left dot is pointing down and spin in the right dot is pointing up.

\subsection{Time evolution of the spin separation process}

Proposed nanodevice is designed in such a way to generate precisely the state  in which spins in the left and the right dots are separated and oriented in opposite direction i.e. like in the $\left|\Psi^\mathrm{sep}_{\pm}\right\rangle$ states. 

  In order to prepare such a spin separated state e.g. the $\left|\Psi^\mathrm{sep}_{+}\right\rangle$ state, we propose a following scheme: Capture two electrons in a single potential well (\(V_1(t_0)=0\)) in the singlet state, then transform the singlet state to linear combination of the singlet and the triplet states by resonantly oscillating RSO coupling and finally separate electrons by raising the potential barrier $V_1(t_0)\rightarrow V_1(t_{sep})=V_0$.

Initially two electrons are confined in the single quantum dot (\(V_1(t_0)=0\)). If one wait long enough, the two electrons will relax to the singlet ground state due to thermalization process\cite{9,10}. 

Let us take a look at the RSO interaction Hamiltonian~(\ref{eq:7}) for two electrons. The central part of the matrix (2nd and 3rd rows and columns) is presented below
\begin{eqnarray}
\begin{pmatrix}
H^{SO}_{22} & 0 \\
0 & H^{SO}_{33}
\end{pmatrix}=
i \alpha(t)\!\begin{pmatrix}
 \partial_{x_1}\!\!-\!\partial_{x_2} & 0 \\
  0 & \partial_{x_2}\!\!-\!\partial_{x_1}
\end{pmatrix}\!.
\label{eq:16}
\end{eqnarray}
It is antisymmetric in the spatial as well as in the spin variables of both electrons, which can be written in a formal way as:
\(H^{SO}\!(x_1,x_2)=- H^{SO}\!(x_2,x_1)\) and \(H^{SO}_{22}=-H^{SO}_{33}\). It has thus symmetries that allow the transition of the system between the singlet and triplet states since the matrix element \(\left\langle T_0\right| H^{SO} \left| S\right\rangle\) is non-zero. 

In order to drive transitions between singlet and triplet state one shall turn on the oscillations of the spin-orbit interaction with a resonant frequency $\omega$ tuned to the energy difference between singlet and triplet state \(\omega=E_{ST}/\hbar\) by applying oscillating voltage to the electrodes $e_1$ and $e_2$.

In simulations we assume sinusoidal alternating electric field with the amplitude of \(5\times10^5\)~V\!/m corresponding to the oscillating voltage with the amplitude of 50~mV applied between the electrodes \(\mathrm{e}_1\) and \(\mathrm{e}_2\) separated by 100 nm distance. The RSO coupling constant is proportional to the electric field \(\alpha_0=\alpha_{3D}e\langle \mathcal{E} \rangle\), and assuming that \(\alpha_{3D} = 5\)~\(\mathrm{nm}^2\),\cite{28,*29} we will obtain \(\alpha(t)=\alpha_0\sin(\omega t)\), where the amplitude of the RSO coupling strength expressed in atomic units is equal to \(\alpha_0\) = 2.5~meV\,nm \(\simeq 0.002\)~\(\mathrm{a_B}\,\mathrm{E_h}\). The $a_B=0.0529$ nm is the Bohr radius and the $E_h=27.211$meV is the Hartree energy.
In order to maintain constant in time $E_{ST}$ (in single dot regime) the electrons are confined in the fixed length InSb section of quantum wire (see Fig. 2), so that the voltages controlling the strength of the RSO interaction coupling do not change the singlet-triplet energy splitting.

\begin{figure}[ht!]
\includegraphics[width=8.7cm]{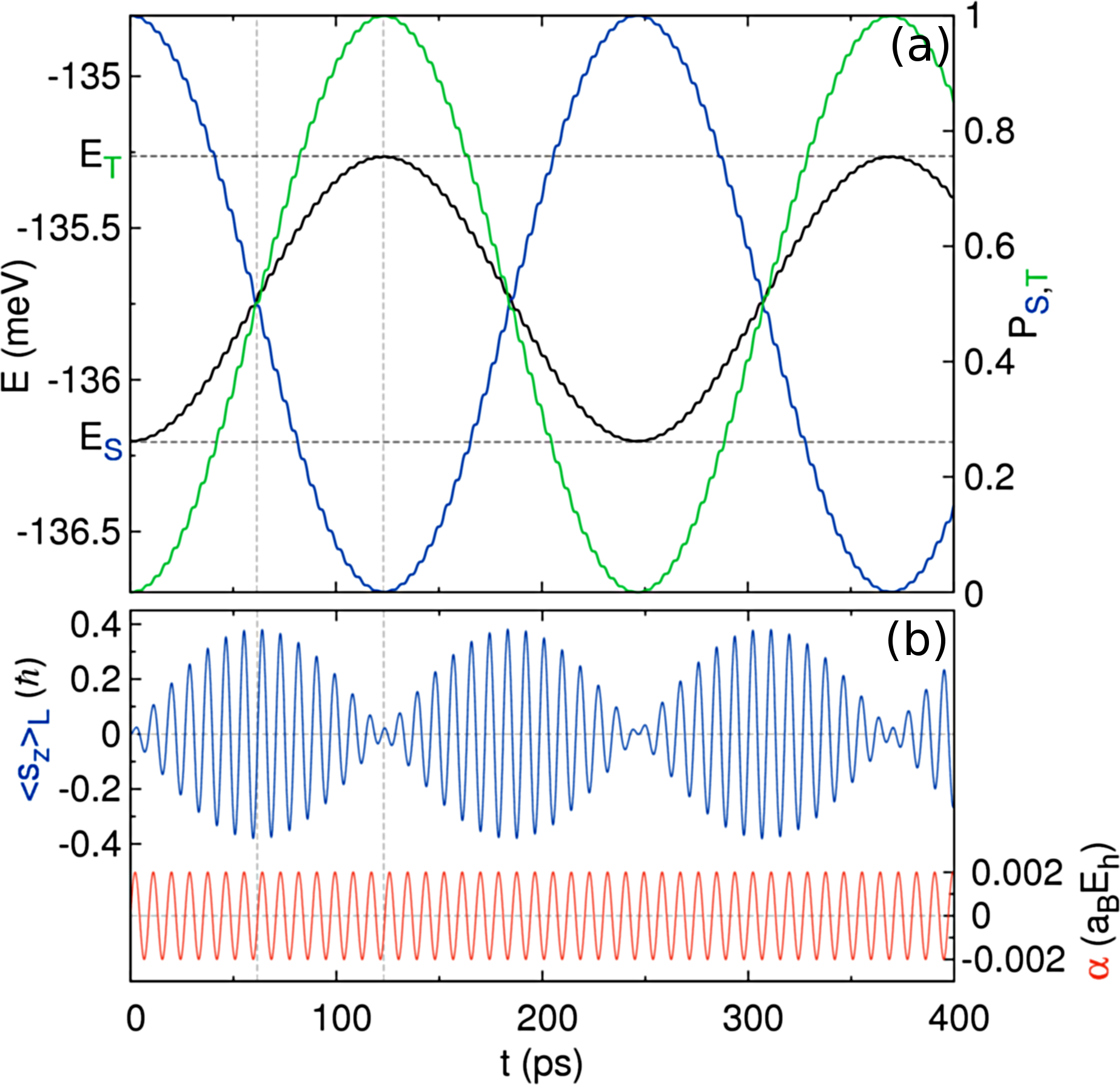}
\caption{\label{fig:5} (color online). The time evolution of the energy of the system E(t) (black curve) and the probability of occupying the singlet state \(P_S(t)\) (blue curve) and the triplet state \(P_T(t)\) (green curve) (a). The evolution of the spin $z$-component in the left half of the InSb nanowire  $\langle s_z(t) \rangle_L$ (blue curve) and the oscillations of the RSO coupling $\alpha(t)$ (red curve) (b).}
\end{figure}

The process of transition from the singlet state to the triplet state induced by resonantly oscillating RSO coupling is presented in Fig.~\ref{fig:5}.
The time evolution of the probability of finding the system in the singlet state is denoted by \(P_S(t)=|\langle S|\Psi(x_1, x_2,t) \rangle|^2\) (blue curve) and the probability of finding it in the triplet state by \(P_T(t)=|\langle T_0|\Psi(x_1, x_2, t) \rangle|^2\) (green curve) in the Fig.~\ref{fig:5}a. At the starting point ($t_0=0$), the system is in the singlet state thus probability \(P_S(t_0)=1\) (\(P_T(t_0)=0\)) and after \(t_T \approx 125\)~ps it drops to zero: \(P_S(t_T)=0\) (rises to one: \(P_T(t_T)=1\)) and then rises (drops) again. The system oscillates between the singlet and triplet state, so we can observe the Rabi oscillations which are characteristic for resonantly driven two-state systems.

The time evolution of the expectation value of the total energy of the system $E(t)=\langle\Psi(x_1,x_2,t)|H|\Psi(x_1,x_2,t)\rangle$ is marked by the black curve in Fig.~\ref{fig:5}a. Its minimum value corresponds to the energy of the singlet state $E(t_0)=E_S$, and its maximum value corresponds to the triplet state energy $E(t_T)=E_T$.

The spin in both dots would only be well-defined in the final state of our simulation, but we want to observe the spin density distribution throughout the entire spin separation process. Thus we introduce an additional parameter - the expectation value of the spin \(z\)-component in the left half of the InSb quantum wire section. It can be calculated as integral over the half-space from the expression that can be defined as the distribution of the spin density of both the electrons:
\begin{align}
\left\langle s_z(t) \right\rangle_L=&\frac{\hbar}{2} \int\limits^{l/2}_0 dx_1 \int\limits^l_0 dx_2 \Big( \Psi^\dag(x_1,x_2,t) \,\sigma_z\!\otimes\!1_2\, \Psi(x_1,x_2,t) \,+ \nonumber \\
&+\Psi^\dag(x_2,x_1,t) \,1_2\!\otimes\!\sigma_z\, \Psi(x_2,x_1,t) \Big),
\label{eq:15}
\end{align}
where \(l\) is the length of the wire. In analogous manner one can calculate $\langle s_z(t) \rangle_R$ - the expectation value of the spin \(z\)-component  in the right half of the nanowire.


\begin{figure}[ht!]
\includegraphics[width=8.7cm]{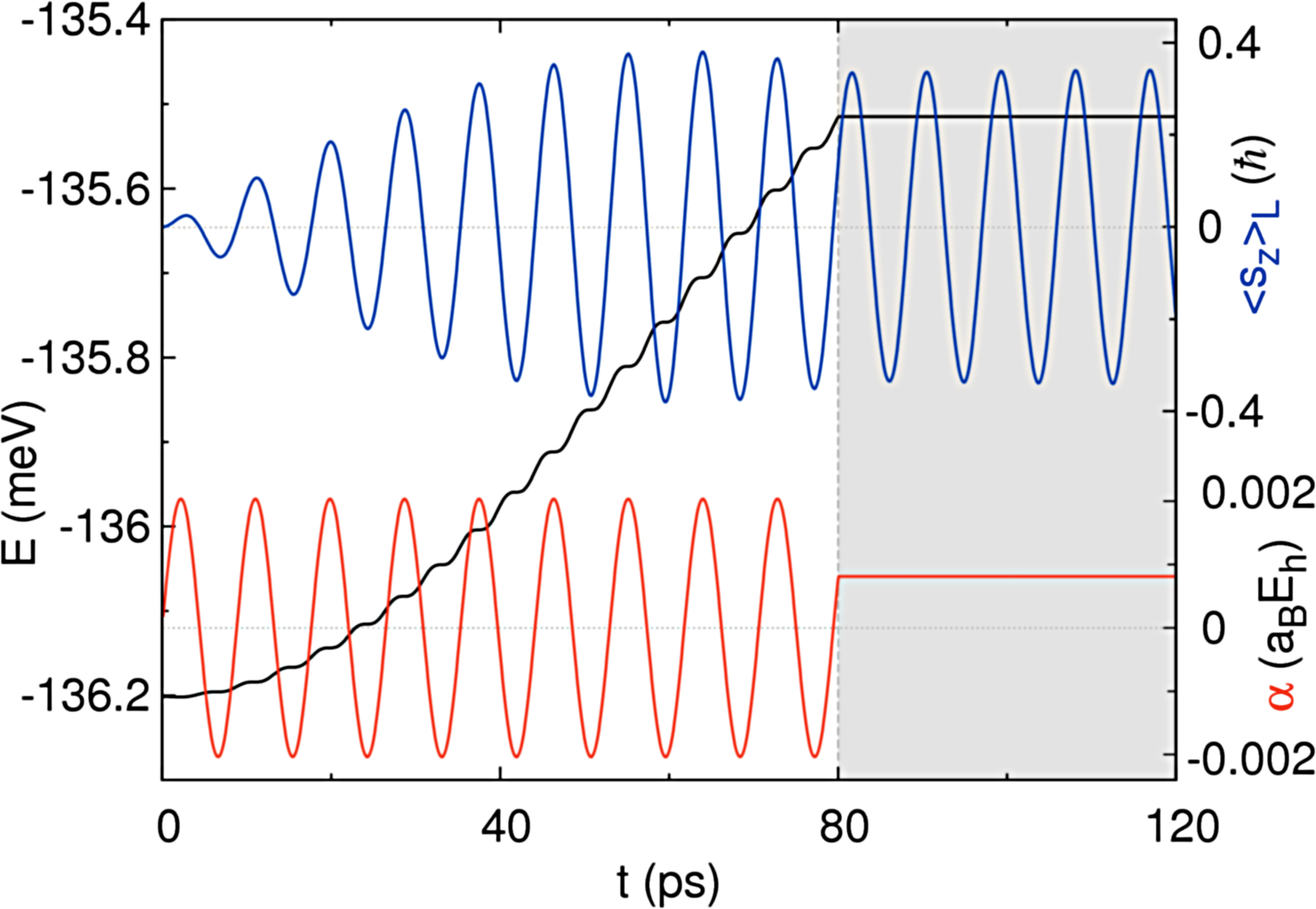}
\caption{\label{fig:6} (color online).
Oscillations of the expectation value of the spin in the left half of the quantum wire $\langle s_z(t) \rangle_L$ (blue curve) induced by the oscillating  Rashba coupling strength $\alpha(t)$ (red curve). The black curve illustrates an increase of the energy of the system driven by the alternating $\alpha(t<80ps)$. It can be seen that when the oscillation of the RSO coupling are stopped the total energy of the system takes finite value $E(t>80ps)=-135.515$meV, as well as the amplitude of the oscillation of the $\langle s_z(t) \rangle_L$.
}
\end{figure}

The state which is a linear combination of the triplet and the singlet states, which correspond to different eigenvalues of the Hamiltonian is not a stationary state. Thus time-varying phase difference between singlet and triplet parts of the wave function causes oscillations of the expectation value of the spin \(\langle s_z(t) \rangle_L\) calculated in this state. Time evolution of the $\langle s_z(t) \rangle_L$ during singlet-triplet transitions can be found on Fig.~\ref{fig:5}b (blue curve). 
The frequency of \(\langle s_z(t) \rangle_L\) (blue curve) is equal to the frequency $\omega$ of the oscillating in time RSO coupling strength $\alpha(t)$ which drives the singlet-triplet transitions (see red curve on the Fig.~\ref{fig:5}b).


 At the starting point, the amplitude of \(\langle s_z(t_0=0) \rangle_L=0\) is zero since only singlet state is occupied. Than it starts to oscillate and amplitude of the \(\langle s_z(t) \rangle_L\) is gradually growing, reaching its maximum value at the moment when the two electron system occupies with equal probability the singlet and the triplet state $P_S(t_e)\approx P_T(t_e)$, which occurs for $t_e\approx60ps$. Then the amplitude of the \(\langle s_z(t) \rangle_L\) is decreasing, and oscillates in time. The periodic modulation of \(\langle s_z(t) \rangle_L\) oscillations are observed.


If we now turn off the oscillations of the RSO interaction coupling (see gray part of the Fig.~\ref{fig:6}), the transitions between the singlet and the triplet states will be terminated.  The oscillations of spin orbit coupling are stopped for $t=80ps$, $\alpha(t>80ps)=const$ (gray part of Fig.~\ref{fig:6}). For $t>80$ps the oscillations of the \(\langle s_z(t) \rangle_L\) (blue curve in Fig.~\ref{fig:6}) will still take place but with a fixed amplitude, because the probability of occupying the singlet and the triplet states is now constant which manifests in the constant energy $E(t>80ps)=const$ (black curve in Fig.~\ref{fig:6}). 
\begin{figure}[ht!]
\includegraphics[width=8.7cm]{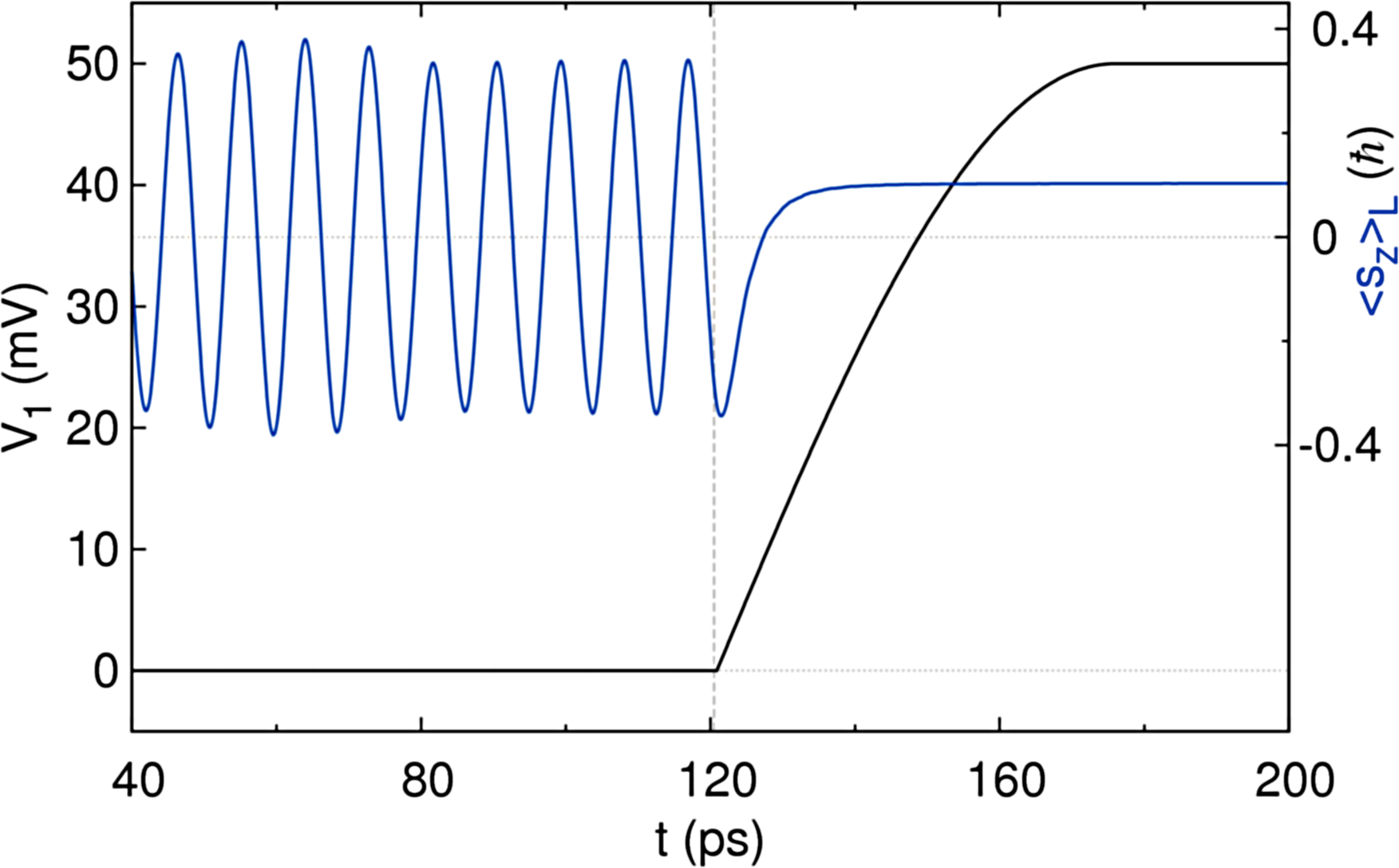}
\caption{\label{fig:7} (color online). Time evolution of the expectation value of the spin in the left part of the nanowire $\langle s_z(t) \rangle_L$ (blue curve) and the potential barrier height \(V_1(t)\) (black curve). For $40ps<t<120ps$ situation corresponds to that, which is depicted in Fig.~\ref{fig:6}. At the moment $t=120ps$ the interdot barrier $V(x,t)$ is started to be lifted up, which stops the oscilation of the spin $\langle s_z(t) \rangle_L$. Finally the spin in the left dot is set to $\langle s_z(t) \rangle_L\approx 0.1\hbar$ value.
}
\end{figure}

Furthermore, oscillations of the expectation value of the spin $\langle s_z(t) \rangle_L$ can be stopped. It can be realized by dividing the section of the quantum wire into two parts separated by a potential barrier. In the simulation below, in order to reach double dot regime, we increase the height of the interdot tunnel barrier in the expression~(\ref{eq:1}) from \(V_1=0\) to \(V_1=V_0\). The process of $\langle s_z(t) \rangle_L$ oscillation shutdown has been presented in Fig.~\ref{fig:7}. We start to raise the barrier for \(t = 120\)~ps which increase the inter-electron distance and in turns we observe the reduction of the $\langle s_z(t) \rangle_L$ oscillations. For the  $t\approx170$ps the interdot barrier reaches its maximal height, and than remains constant $V_1(t>170ps)=V_0$. 
The height of the barrier $V_1=V_0$ has been chosen so that at the final point the electrons can be separated by the impenetrable barrier and thus the energy of the singlet and the triplet states is equal as well the exchange interaction is reduced to zero. The oscillations of \(\langle s_z(t) \rangle_L\) are stopped.
In this case the value of \(\langle s_z(t) \rangle_L\) has been set at the level of 0.1~\(\hbar\). In this example moment of the turning off the oscillations of RSO coupling strength as well as the moment when the barrier is started to be raised up has been chosen arbitrarily.

\begin{figure}[ht!]
\includegraphics[width=8.7cm]{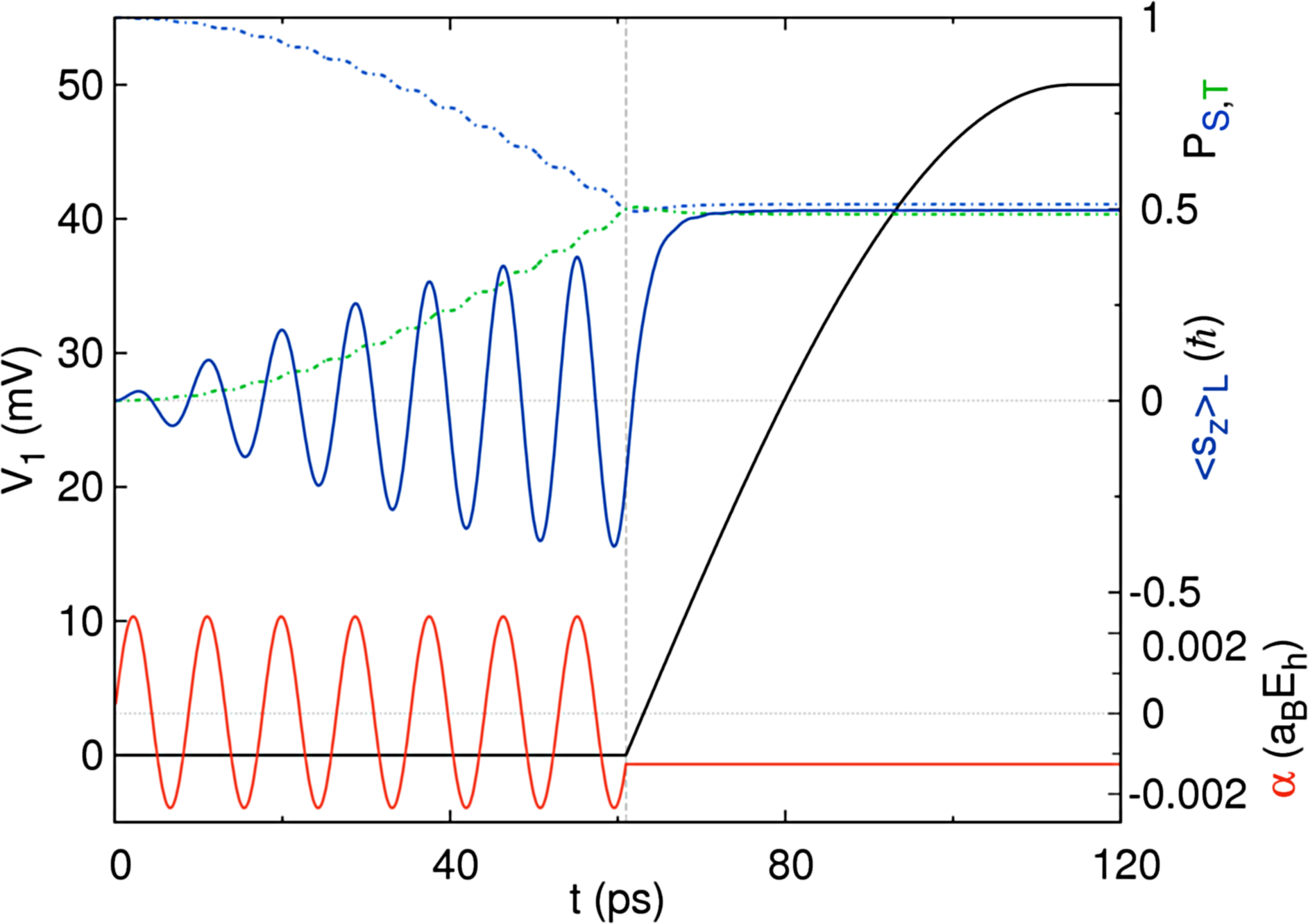}
\caption{\label{fig:8} (color online). The time evolution of the two-electron system (same as in Fig.~\ref{fig:6} and in Fig.~\ref{fig:7}) with carefully adjusted moment of the Rashba coupling oscillation shutdown and the time of the barrier formation. The dash blue (green) curve shows the time evolution of the probability of occupation the singlet \(P_S(t)\) (triplet \(P_T(t)\)) state. The red solid curve shows the oscillations of the RSO coupling, the blue solid curve shows the course of the spin's expectation value in the left dot, and the black solid curve presents the evolution of the potential barrier height $V_1(t)$ that separates the quantum wire into two parts.}
\end{figure}

However our intention is to split up electrons with opposite spins and obtain one of the $\left| \Psi^\mathrm{sep}_{+}\right\rangle$ or $\left| \Psi^\mathrm{sep}_{-}\right\rangle$ states for which spin in the left (right) dot is respectively oriented up (down) or down (up). This can be achieved by carefully matching the moment $t_{off}$ when the oscillations of spin-orbit coupling strength are stopped and the time of the interdot barrier formation. An example which illustrates time evolution of preparation 
the system in the $\left| \Psi^\mathrm{sep}_{+}\right\rangle$ is presented in Fig.~\ref{fig:8}. Oscillations of the Rashba coupling are turned off when the wave function of the two electron system becomes a balanced linear combination of the singlet and the triplet state, which happens when the dashed blue and green curves reach the value of $P_S(t_{off})=P_T(t_{off})=0.5$. This takes place at \(t_{off} = 60\)~ps. At the same time $t_{off}$, we start lifting the barrier dividing the wire into two halves and split up both electrons. The duration of barrier formation $t_{form}=t_{sep}-t_{off}$ must be such that, in the final moment - $t_{sep}$, the phases of the singlet and triplet parts of the wave function are equal, like in the expression~(\ref{eq:14}), defining the $\left| \Psi^\mathrm{sep}_{+}\right\rangle$ state. The spins of the electrons are then separated. The left dot will contain the electron with spin up (the spin projection on \(z\)-axis in the left dot is equal to $\langle s_z(t_{sep}) \rangle_L=\hbar/2$), and in the right dot the spin will be opposite and equal to $\langle s_z(t_{sep}) \rangle_R=-\hbar/2$.

\begin{figure}[ht!]
\includegraphics[width=8.7cm]{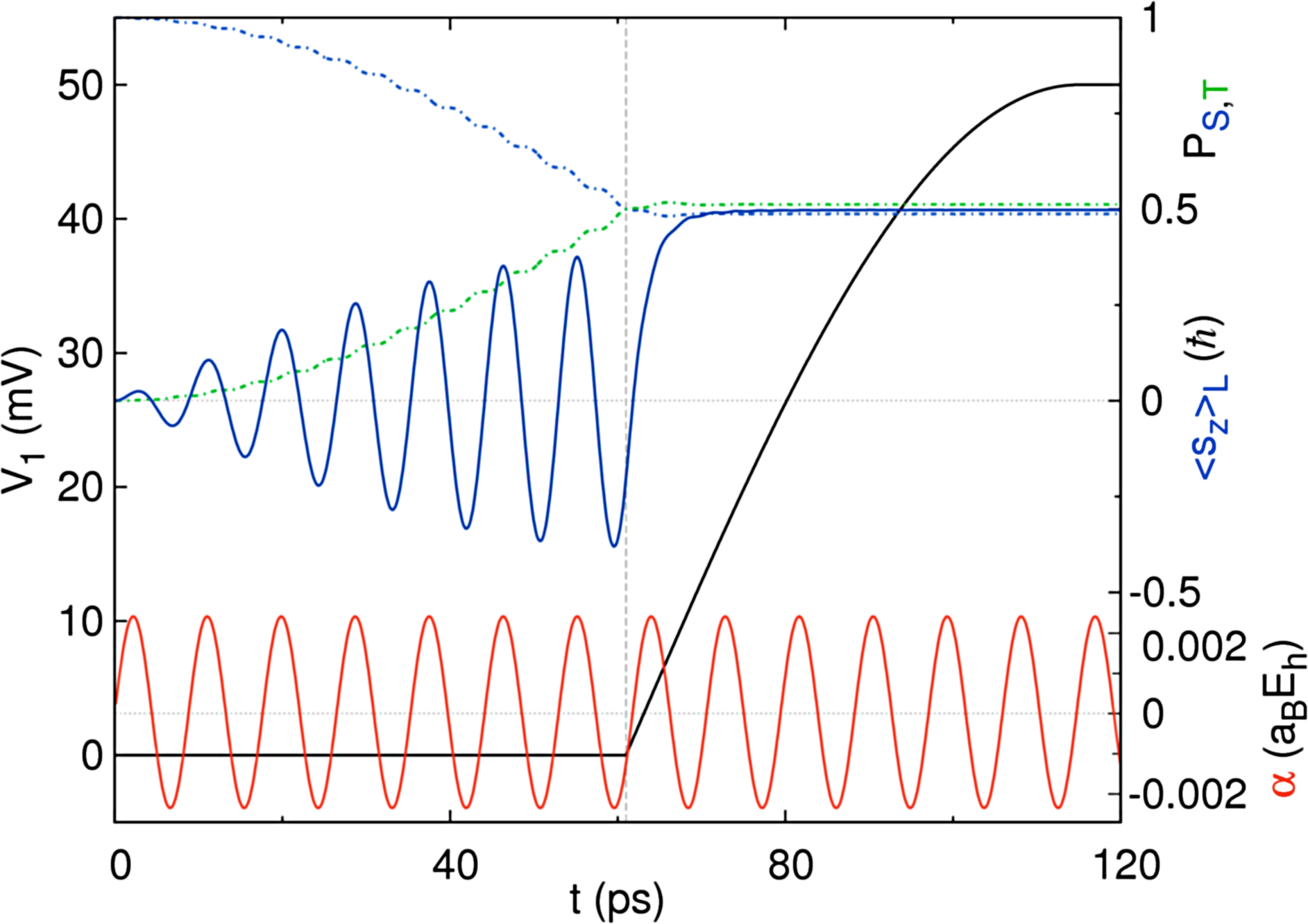}
\caption{\label{fig:9} (color online). Same as in Fig.~\ref{fig:8} but in this case we do not turn off the oscillations of the RSO coupling, $\alpha(t)$ is oscillating all the time.}
\end{figure}

It is also possible to perform a different, slightly simplified version of the spin separation process. The results are shown in Fig.~\ref{fig:9}. In this approach we do not stop the oscillation of the RSO coupling strength, $\alpha(t)$ is oscillating all the time. All we need to do is to raise the potential barrier $V_1\rightarrow V_0$ at the right moment. The lifting barrier will reduce the energy difference between the triplet and singlet states.  Due to the strongly resonant nature of singlet triplet transitions driven by the oscillating RSO coupling strength, the singlet triplet transitions as well as the oscillations of the $\langle s_z(t) \rangle_L$ are terminated automatically when the barrier is formed.
\begin{figure}[ht!]
\includegraphics[width=8.7cm]{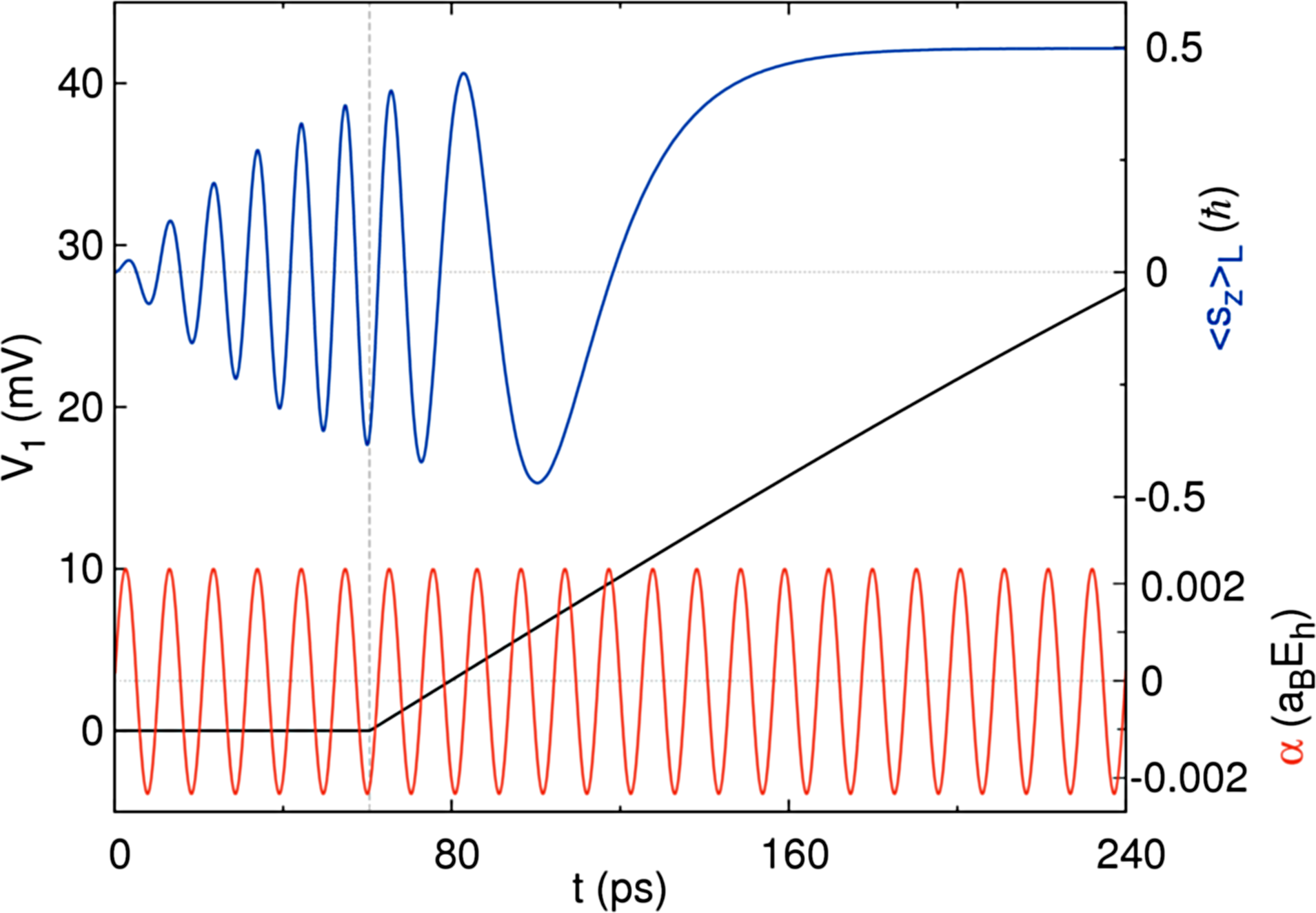}
\caption{\label{fig:10} (color online). Same as in Fig.~\ref{fig:9} but in this case the barrier is lifted much slower: aproximately 3 times slower.}
\end{figure}


However the way in which the barrier is raised is not significant: there are not restrictions on speed of raising of the barrier. Even a relatively rapid increase of the barrier does not disturb the stability of the wave function. The moment when the barrier completely removes the exchange interaction is  important. The spin of the electron in the left dot should at this time reach a  value of $\hbar/2$. The best approach will be to adjust this moment experimentally. To make it easier, the barrier lifting can be slowed in the final phase. The time evolution of the system with such improvement is added in Fig.~10.

In the paper we presented simulation of the nanodevices which is able to generate $|\Psi^{sep}_+\rangle$ state. However one can obtain the another spin separated state $|\Psi^{sep}_-\rangle$ with the spin down (up) in the left (right) dot. In order to generate such a state one has to follow analogous procedure as presented above with that difference, that the oscillation of the RSO coupling strength should have a different sign: $\alpha(t)=-\alpha_0\sin(\omega t)$.

Proposed device allows for high fidelity initialization of the electron spin in the spin up or the spin down state e.g. for the quantum computation purposes.
We made its simulations for InSb material parameters, however proposed electron spin separation scheme is universal and can be realized for other materials.
The difference will be in the singlet triplet splitting energy $E_{ST}$ and thus frequency $\omega$ of oscillations of the RSO coupling. In the presented model we chose parameters for which singlet-triplet splitting energy is large which results in spin precession frequency of the order of 100GHz. We are aware that such fast oscillation may be challenging during experimental implementation however we take these parameters in order to check the limit of the adiabaticity of the process. We have showed that for such parameters spin separation process is adiabatic. Thus any slower process will be also adiabatic. One can slower the process by keeping non zero barrier high or take the longer section of InSb wire. It will reduce $E_{ST}$ energy and thus lower the frequency $\omega$ of RSO coupling strength oscillations. In this case the proposed spin separation scheme is simply more resistant to mismatches in the barrier switching times.





\section{SUMMARY AND CONCLUSIONS}
In conclusion, we have proposed nanodevice - gated semiconductor nanowire -  that allows for separating spins of two electron system, and consequently prepare electron spin i.e. in the well define spin up or spin down state.
In order to generate such a state the following procedure is applied. 
Electrons are initially confined in the single quantum dot formed in the InSb nanowire in the singlet state. Then the spin-orbit (Rashba) interaction, with the  oscillating amplitude with a frequency equal to the energy difference between the triplet and singlet states, gradually transfers the two electrons system to to an excited state. Once the system is in a state that is a balanced linear combination of the singlet and the triplet state, by using an electrostatically generated potential barrier, the quantum dot is divided into two parts (double dot regime), in which each electron is confined. A properly selected rate of barrier formation will allow us to separate the spins of the two electrons, so that e.g. the electron spin in the left dot is up, and the one in the right dot is down. Thus proposed device allows for preparation an electron in the spin up or spin down state with very high precision.


The proposed nanodevice is all-electrically controlled (by the voltages applied to the electrodes) without the need of application of magnetic field at any step of the spin separation process. Thus it can address individual electron spin qubits in selective manner which can be useful for scalability purposes.
The nanodevices operates within tens of picoseconds which is much less than typical decoherence time for electron spin qubits in these type of materials.

\begin{acknowledgments}
J.~P. has been partly supported by the EU Human Capital Operation Program, Polish Project No. POKL.04.0101-00-434/08-00.
P.~S. was supported by the Polish National Science Center (Grant No. DEC-2011/03/N/ST3/02963), as well as by the "Krakow Interdisciplinary PhD-Project in Nanoscience and Advanced Nanostructures" operated within the Foundation for Polish Science MPD Programme, co-financed by the
European Regional Development Fund.
\end{acknowledgments}

\nocite{*}

\bibliography{BJ12486_res} 

\begin{thebibliography}{10}
\bibitem{comm}%
  \BibitemOpen
  \bibinfo {note} {It is important to note that the electric field
  $\mathcal{E}_z$, generated by the electrode $e_3$ when the barrier is lifted
  up, is another source of the RSO interaction: $H_R^{z}= \alpha_zk_x\sigma_y$.
  However $\alpha_z=\alpha_{3D}e\langle\mathcal{E}_z\rangle\approx1.5$ meVnm
  (for voltage applied to the $e_3$ of the order of $50$ mV) gives spin-orbit
  length equal to $\lambda_{SO}^z=\hbar^2/(m^*\alpha_z)\approx 3600$ nm. During
  lifting the barrier electrons move only by the 25 nm distance which is much
  shorter than $\lambda_{SO}^z$, thus one can expect that contribution of the
  $H_R^{z}$ will be negligible. Electrons experience the $H_R^{z}$ interaction
  only when the barrier is nonzero (non zero voltage applied to the $e_3$) and
  they posses nonzero momentum $k_x$. We made an explicit simulation including
  $H_R^{z}$ in the total Hamiltonian (1) and as one may expect it does not
  influence noticeably any results present in the paper. Transitions to the
  $T_+$ and $T_{-}$ states are negligibly small.}%
  \bibAnnoteFile{Stop}{comm}%
\bibitem{comm0}%
  \BibitemOpen
  \bibinfo {note} {In numerical calculations we used a treshold $E_{ST}<
  1\mu{}eV$}%
  \bibAnnoteFile{NoStop}{comm0}%
\end{thebibliography}

\end{document}